\documentstyle[psfig,twoside]{article}
\begin{document}\hbadness=10000\thispagestyle{empty}
\pagestyle{myheadings}\markboth{Henrik Bohr, Patrick McGuire,
Chris Pershing and Johann Rafelski}{Threshold Noise as 
Source of Volatility in RSANN}
\title{THRESHOLD NOISE AS A SOURCE OF VOLATILITY IN\\ 
RANDOM SYNCHRONOUS ASYMMETRIC NEURAL NETWORKS} 
\author{Henrik Bohr\thanks{Center for Biological Sequence Analysis, 
The Technical University of Denmark, Building 206, DK-2800 Lyngby, Denmark,
EMAIL: hbohr@cbs.dtu.dk}, 
Patrick McGuire\thanks{Center for Astronomical Adaptive Optics, 
Steward Observatory, University of Arizona, Tucson AZ 85721,
EMAIL: mcguire@as.arizona.edu}, 
Chris Pershing\thanks{EMAIL: chris@physics.arizona.edu}\,
and Johann Rafelski\thanks{EMAIL: rafelski@physics.arizona.edu}\\$\ $\\
Department of Physics, University of Arizona, Tucson AZ 85721}
\date{November 24, 1997}
\maketitle
\begin{abstract}
We study the diversity of complex spatio-temporal patterns of
random synchronous asymmetric neural networks (RSANNs). Specifically, we
investigate the impact of noisy thresholds on network performance and find
that there is a narrow and interesting region of noise parameters where
RSANNs display specific features of behavior desired for rapidly `thinking'
systems: accessibility to a large set of
distinct, complex patterns.
\end{abstract} 
\section{Introduction}
Random Synchronous Asymmetric Neural Networks (RSANNs) with fixed synaptic coupling strengths and fixed
neuronal thresholds have been found to have access to a very limited set of
different limit-cycles (Clark, K\"urten \& Rafelski, 1988; Littlewort,
Clark \& Rafelski, 1988; Hasan, 1989; Rand, Cohen \& Holmes, 1988).
We will show here, however, that
when we add a small amount of temporal random noise by choosing the neural
thresholds not to be fixed but to vary within a narrow gaussian
distribution at each time step, we can cause transitions between different
quasi-limit-cycle attractors. 
Shifting the value of neuronal thresholds randomly
from a gaussian distribution which varies on a time scale much slower than
the fast synaptic dynamics, we control the timing of the transition
between limit-cycle attractors. Hence we can gain controllable and as
we believe biologically
motivated access to a wide variety of limit-cycles, each displaying 
dynamical participation by many neurons.

Perfect limit-cycles in which the network returns to some earlier state do
not exist in real biological systems, due to many complicating factors:
membrane potential noise (Little, 1974; Clark, 1990; Faure, 1997), the complexity of
biological neurons, the continuously-valued signal transmission times
between neurons, and the lack of a clock to synchronously update all neurons,
to mention a few.
Therefore the case of {\it approximate} limit-cycles is  more
biologically reasonable.
Indeed, the appearance of limit-cycle behavior in
central pattern generators is evidence for such temporal behavior in biological
 systems (Hasan, 1989; Marder \& Hooper, 1985). We also believe that the biologically significant limit-cycles 
are those in which a great fraction of the neurons actively
participate in the dynamics. 

If these limit-cycles represent  brain activity states which we could call `thoughts',
and if a sequence of spontaneous
or externally controlled transitions from one limit-cycle to another
limit-cycle represents a `reasoning process', then neural network systems which
exhibit a diversity of accessible limit-cycles (thoughts) with capability to
move rapidly between them could also exhibit a large
variety of different `trains of thought'. A system which can access {\it many%
} limit-cycles should always be able to access a novel mode; hence the system
would have the potential to be a {\it creative} system. Our current work thus
demonstrates conditions sufficient to allow access to creative dynamical behavior.

In Section 2 we introduce RSANNs along with the concept of threshold noise,
as well as the details of an RSANN's implementation. In Section 3, we provide
the algorithm that contrasts limit-cycles, which is of prime
importance in our work.  
In our quantitative investigations we need to introduce with more precision
concepts which intuitively are easy to grasp, but which mathematically are
somewhat difficult to quantify. We define {\bf eligibility} in Section \ref
{elig} as an entropy-like measure of the fraction of neurons which actively
participate in the dynamics of a limit-cycle. In order to quantify the
RSANN's accessibility to multiple limit-cycle attractors, we define 
{\bf diversity} in Section 4.2 as another entropy-like measure,
calculated from the probabilities that the RSANN converges to each
of the different limit-cycles.  As a measure of the creative
potential of a system, we introduce the concept of {\bf volatility} as the
ability to switch from one particular highly-eligible cyclic mode to many
other highly-eligible cyclic modes. We mathematically define volatility in
Section \ref{volatile} to be an entropy-like measure of the number of ${\em 
different}$ limit-cycle patterns easily available to the net, weighted by
the eligibility of each limit-cycle attractor. We find that in terms of
these variables, as the neuronal threshold noise, $\epsilon $, increases,
our RSANN exhibits a phase transformation at $\epsilon =\epsilon_1$ from a small
number to a large number of different ${\em accessible}$ limit-cycle
attractors (Section 4.2), and another phase transformation at 
$\epsilon =\epsilon _2>\epsilon_1$ from high eligibility to low eligibility
 (Section 4.1). Our main result is that the
volatility is high only in presence of threshold noise of suitable fine-tuned 
strength chosen between $\epsilon _1\leq \epsilon \leq \epsilon _2$, so
allowing access to a diversity of eligible limit-cycle attractors (Section 4.3).

\medskip\ 

\section{\bf Random asymmetric neural networks with threshold noise} 

Random asymmetric neural networks (RSANNs)
(Bressloff \& Taylor, 1989; Clark, Rafelski \& Winston, 1985; Clark, 1991), with $w_{ij}\ne w_{ji}$, differ from 
symmetric neural networks (SNNs) (Hopfield, 1982),
and offer considerably more biological realism, since real neural
signals are unidirectional. Due to the lack of synaptic symmetry,
RSANNs have a non-simple behavior with different limit-cycle attractors with
possible period lengths\footnote{$\Delta t$
is the time necessary for a neural signal to propagate from one
neuron to the next and is assumed to be identical for all synapses (McCullough \& Pitts, 1943);
hereafter, we will take as the unit of time $\Delta t = 1$, so that a limit-cycle of period $L$
will be $L$ time steps long} $L \Delta t\gg\Delta t$,
even with extremely long periods (greater
than $10,000$ time steps) when the neuronal thresholds have been finely tuned
(Clark, K\"urten \& Rafelski, 1988; McGuire, Littlewort \& Rafelski, 1991;
McGuire {\it et al.}, 1992).

Consider a net of $N$ neurons with firing states which take binary values
(i.e. $a_i\in {0,1}$). Each neuron is connected to M other pre-synaptic
neurons by unidirectional synaptic weights. A firing pre-synaptic neuron $j$
will enhance the post-synaptic-potential (PSP) of neuron $i$ by an amount
equal to the connection strength, $w_{ij}$. Inhibitory neurons (chosen with
probability $I$) have negative connection strengths.

If the $i$th neuron's PSP, $c_i$, is greater than its threshold, $V_i$, then
it fires an action potential: $a_i\rightarrow 1$. We parametrize the
thresholds, $V_i$, in terms of `normal' thresholds, $V_i^0$, a mean bias
level, $\mu ,$ and a multiplicative threshold-noise parameter, $\beta _i$: 
\begin{equation}
\label{ThreshNoise1} 
\begin{array}{ccc}
V_i(\beta _i) & = & (\mu +\beta _i)V_i^0 \\ 
V_i^0 & = & \frac 12\sum\limits_{k=1}^Mw_{ik} 
\end{array}
, 
\end{equation}
If $\mu =>1\,$and $\beta _i=>0$, we recover normal thresholds ($V_i => V_i^0$).

\begin{figure}[t]
\centerline{\hbox{
\psfig{figure=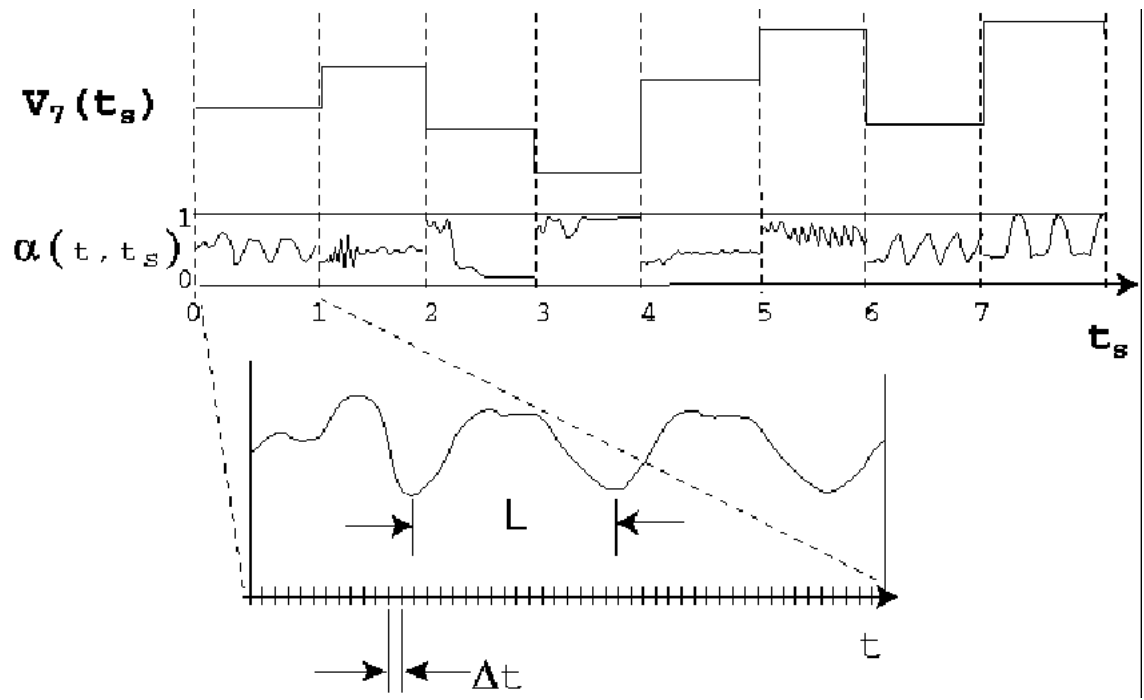,width=15cm}
}}
\caption[Timing Diagram]{A qualitative sketch showing the threshold for neuron \#7 ($V_7(t_s)$)
and the spatially-averaged firing rate, $\alpha(t,t_s)$
as a function of slow-time ($t_s$) and in the inset the $\alpha(t,t_s)$
as a function of fast-time $t$.
\label{Timing}}
\end{figure}

In the presence of even a small amplitude of neural-threshold noise 
effective on the same time-scale as the transmission-time of neural
impulses from one neuron to the next neuron, the neural net will never
stabilize into a single limit-cycle attractor. Rather, if the noise
amplitude is not too high, the net will continually make transitions
from one almost-limit-cycle to another almost-limit-cycle\footnote{the prefix
`almost' here, permits occasional misfirings}. This `non-stationary'
characteristic is common to any volatile system, but makes computer
simulation and characterization difficult because one never knows
{\em a priori} when the neural network will `jump' to 
a new attractor basin. In order to maintain stability and
make simulation easier, but also for reasonjs of biological
reality we will give the neural threshold noise a much
longer time scale, $t_s$, than the shorter signal-transmission/neural-update
time scale, $\Delta t$. Consequently, for each discrete slow-time threshold-update
step, $t_s$, there will be several hundred to several thousand fast-time
neural-update steps, $\Delta t$, during which the thresholds $V_i(\beta _i(t_s))$
are `quenched' (or do not change) and the neural states $a_i(t)$ are allowed
to change via Eq. \ref{DynamEq}. Only at the next slow-time step are the
thresholds allowed to change,
but then kept quenched again during the many fast-time neural-update steps, see Figure \ref{Timing}. We see here the evolution of a threshold value and
neural activityi on slow-time, $t_s$, while in the insert the fine-structure of limit-cycle dynamics as a function of fast-time is displayed.

For a given limit-cycle search, labelled by the slow-time-scale parameter $t_s$,
the neurons initially possess random firing states; 
where the fraction of initially firing neurons
is also chosen randomly to be between $0$ and $1$.
 At a given slow-time step, $t_s,$ we initialize the neuronal
threshold parameters $\beta_i$ with zero-mean gaussian noise of width,
$\epsilon$, 
where each $\beta _i$ is chosen independently for all $i$.

This `slow threshold noise' makes biological sense because
the concentrations of different chemicals in the brain changes
on a time-scale of seconds or even minutes, thereby changing the
effective thresholds of individual neurons on a time
scale much longer than the update time, $\Delta t$ (milliseconds).
Some biologists therefore believe that neural
thresholds are `constant' and noiseless (Fitzhugh-Nagumo Model
(Pei, Bachmann \& Moss, 1995));
others unequivocably feel that neurons live in a very noisy environment,
both chemically and electrically (Little, 1974); a model which
includes slow threshold noise varying randomly by $O(10^{-3}$ seems
consistent with both points of view -- this will be the model developed
here. 

Thereafter the firing state $a_i(t,t_s)$ of the $i$th neuron as a function
of the rapidly-updating time parameter $t$ is given by the following equations: 
\begin{equation}
\label{DynamEq} 
\begin{array}{ccccc}
a_i(t,t_s) & = & \theta (c_i(t,t_s)-V_i(t_s)) &  & \mbox{(a)} \\ c_i(t,t_s)
& = & \sum\limits_{j=1}^Mw_{ij}a_j(t- \Delta t,t_s) &  & \mbox{(b)} \\ 
V_i(t_s) & = & (\mu +\beta _i(t_s))V_i^0\,\,\,\,\,\,, &  & \mbox{(c)} 
\end{array}
\end{equation}
where $\theta $ is the step-function.

\subsection{\bf Implementation}

The prescription given above of an RSANN is implemented in a computer
program on a fast workstation.
We update all neuron firing states in parallel, or `synchronously', as
opposed to serial, or `asynchronous', updating in which only one neuron
or a small group of neurons is
updated at a given time step.
The connection strengths $w_{ij}$ are integers chosen
randomly from a uniform distribution, $\left| W_{ij}\right|
\in (0,50000]$. We have performed a cross-check simulation in which the
connection strengths are double-precision real numbers, and have found no
difference between the integer-valued connection-strength simulations and
the real-valued connection-strength simulations. The
normal thresholds are double-precision numbers. We chose $K$ incoming
connections per neuron, where $K=10$. The fraction of inhibitory connections
is randomly chosen, here with probability, $I=0.3$.
We have studied networks with 
$N$ neurons, where $N\in \{10,20,30,40,50,100\}.$ The network size is
primarily constrained by the extremely long limit-cycles
or transients which get longer for larger networks, especially when
the thresholds are near-`normal', as given by $V_i^0$ in Eq. \ref{ThreshNoise1}.

\smallskip\ 

\subsection{{\bf Updating and Limit-Cycle Search
\label{Updating}}}

We simulate slow threshold noise by reinitializing the thresholds via
equation \ref{ThreshNoise1}, at each slow-time step 
$t_s$. Initially, during this slow-time step $t_s$, the neural firing states
are updated by the dynamical equations \ref{DynamEq} for $W_n=W_0=128$
fast-time steps, $t.$ After the $W_n$ fast-time updates, an iterative
computer routine carefully scans the record of the spatially-averaged firing
rate, $\alpha (t,t_s)$, for $t\in [0,W_n],$ to check for exact periodicity:
\begin{equation}
\label{activity} 
\begin{array}{ccccc}
\alpha (t,t_s) & \equiv & \frac 1N\sum\limits_{i=1}^Na_i(t,t_s) & & \mbox{(a)}  \\ 
\alpha (t+L,t_s) & = & \alpha (t,t_s), & \forall t\in [0,W_n],  & \mbox{(b)}
\end{array}
\end{equation}
where the limit-cycle period is is identified to be $L.$ The iterative routine
carefully scans each candidate limit-cycle to ensure perfect periodicity of 
$\alpha (t,t_s)$ for at least $P$ periods (typically $2$ or $4$ periods)
of the limit-cycle. If the scanning routine finds that the
spatially-averaged firing rate has converged to a limit-cycle (with 
$L<W_n/P$, see Figure \ref{finger}), then we halt the fast-time updating, `record' the limit-cycle,
increment the slow-time ($t_s=t_s+1$), give the thresholds new initial
conditions by choosing new $\beta_i$ in Eq. \ref{ThreshNoise1} and the 
neuronal firing states new initial conditions,
and begin a new limit-cycle search. On the other hand, 
if the scanning routine finds that the
spatially-averaged firing rate has not yet converged to a limit-cycle, we then
let $W_{n+1}=2W_n$, and update the network for another $W_{n+1}$
fast-time steps. After the updating, the iterative limit-cycle scanning
routine searches again for at least $P$ periods of exact periodicity of 
$\alpha (t,t_s)$, but only within the new time-window, $t\in [W_n,W_{n+1}]$.
We repeat this limit-cycle search algorithm until we find a limit-cycle,
or until $W_n=W_{\max }$ (where $W_{\max }$ is typically $4096$ or $8192$ time
steps), whichever comes first.

If the limit-cycle period is much shorter than the width of the time-window, 
\begin{equation}
L<<\Delta W\equiv W_{n+1}-W_n, 
\end{equation}
we will observe many more repetitions of the limit-cycle than the requisite
$P$. If no limit-cycle is observed during the limit-cycle search, then the
RSANN either has a very long transient or a very long period limit-cycle,
see Clark (1991) or Littlewort, Clark \& Rafelski (1988)  for a discussion of the correlation 
between transient-length and limit-cycle period.

\smallskip 

\section{Limit-cycle comparison}

From the fast-time-averaged single neural firing rates and from
estimates of each
neuron's neural fast-time variance during a limit-cycle, we can compare two
limit-cycles found at two different slow-time steps by a chi-square measure
of the difference between two simulated distributions.

We will estimate the likelihood that two different limit-cycles (labelled by
the slow-time indices $t_s$ and $t_s^{\prime }$) belong to different basins
of attraction, by comparing the time-averaged single-neuron firing rate
spatial vectors, $A_i(t_s)$ and $A_i(t_s^{\prime })$, where the subscript $i$
is the spatial index of the neuron. Often, limit-cycles
with different periods $L(t_s)$ and $L(t_s^{\prime })$ will be remarkably
similar, with only an occasional slight difference between the fast-time
recordings of the spatially-averaged firing rates $\alpha (t,t_s)$ and $\alpha (t,t_s^{\prime })$ (see eq. \ref{activity}), caused by an occasional neuron misfiring:

\begin{equation}
\left| \Delta \alpha (t)\right| \equiv \left| \alpha (t,t_s)-\alpha
(t,t_s^{\prime })\right| \sim O(\frac 1N). 
\end{equation}
Conversely, limit-cycles with the same period, $L(t_s)=L(t_s^{\prime })$,
will often have grossly different $A_i(t_s)$ and $A_i(t_s^{\prime })$,
especially for short-period limit-cycles.

\subsection{\bf Estimate of neuronal time-variance}

\begin{figure}[t]
\centerline{\hbox{
\psfig{figure=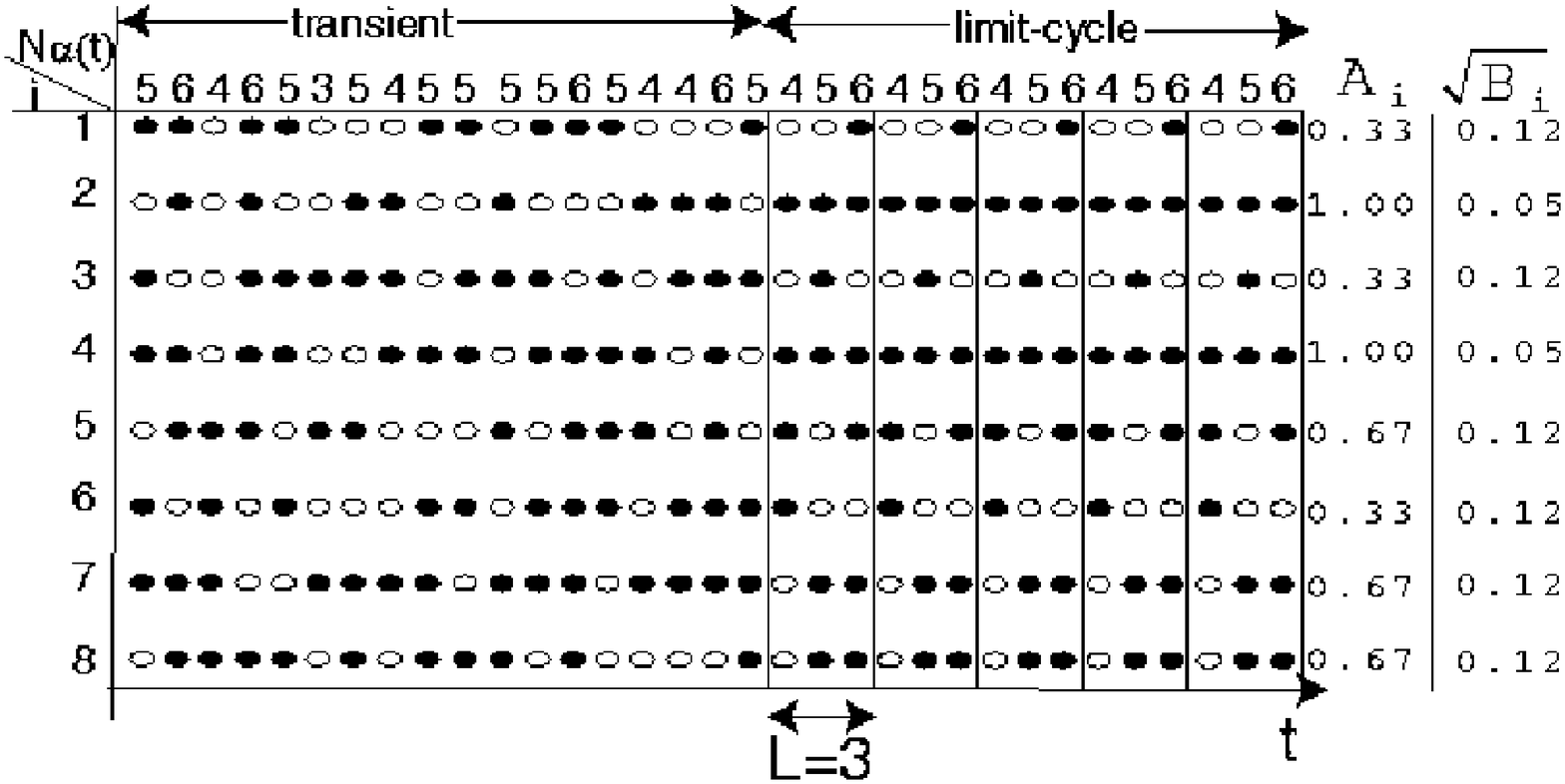,height=4.0in,width=15cm}
}}
\caption[Fingerprint]{A diagram showing a transient of length 18 fast-time steps
and a limit-cycle of period $L=3$ for $\Delta W = 15$ fast-time steps
and for N=8 neurons. A filled-in circle represents $a_i=1$, and an empty
circle represents $a_i=0$.
The limit-cycle period is determined by searching for periodicity
in $N\alpha(t)$. A `fingerprint' of the limit-cycle is then taken
by measuring the time-averaged single-neuron firing rates, $A_i$,
and also measuring the variance from the mean $B_i$ for each of the neurons.
Values of $N\alpha(t)$, $A_i$ and $\sqrt{B_i}$ appropriate for this example are
also shown in the figure.
\label{finger}}
\end{figure}

We `record' each limit-cycle's period, $L(t_s)$, and its time-averaged
single-neuron firing rates, $A_i(t_s)$:

\begin{equation}
A_i(t_s) \equiv \langle a_i \rangle _t \equiv 
\frac 1{\Delta W}\sum\limits_{t=1+W_n}^{W_n+\Delta W}a_i(t,t_s), 
\end{equation}
for all neurons $i$ (see Figure \ref{finger}). Since $a_i^2=a_i$ for $a_i\in \{0,1\}$,
we have $\langle a_i^2 \rangle = \langle a_i \rangle$, and we
find that the variance (of a single measurement from the mean), $b_i(t_s)$, will be:

\begin{equation}
b_i(t_s) \equiv  \langle (a_i - \langle a_i \rangle _t ) ^2 \rangle _t
 = \langle a_i^2(t,t_s)\rangle _t-\langle a_i(t,t_s)\rangle _t^2   
 =  A_i-A_i^2 \, .
\label{Variance}
\end{equation}
For $A_i  \sim 0.5 $, we find that the standard deviation of a single measurement
of the firing state from the mean firing rate is $\sqrt{b_i} \sim 0.5$. However,
for different limit-cycle attractors, the difference between firing rates of
a particular neuron $i$ is $|A_i(t_s) - A_i(t_s^{\prime})| \sim 0.05 \ll 0.5$.
Fortunately, we measure the firing state of each neuron many times ($\Delta W >> 1$) to
determine the mean firing rate, so we can also estimate the variance of the mean, $B_i$, as: 

\begin{equation}
B_i = b_i/\Delta W.
\label{Variance2}
\end{equation}
For $\Delta W \sim 1000$ and $A_i \sim 0.5$, this
gives a standard deviation of the mean of $\sqrt{B_i} \sim 0.01$, which
is a little smaller than the variation seen between different limit-cycle
attractors (see Figure \ref{finger}).  Therefore, we choose to use the variance of the mean, $B_i$,
rather than the variance of a single measurement from the mean for our
limit-cycle comparison tests. If we choose a longer observation time, $\Delta W$,
then we can better discriminate between different limit-cycles by using $\chi^2$
test discussed below.

Since the variance $B_i$ is $0$ for $A_i=0$ or $A_i=1$ (likely when the RSANN
has low eligibility), a $\chi ^2$ comparison of $A_i(t_s)$ and 
$A_i(t_s^{\prime })$ will be plagued by division-by-zero problems. In this
case, a maximum likelihood comparison using Poisson statistics (Smith,
Hersman \& Zondervan, 1993), or
maybe the Kolmogorov-Smirnov test
would be more appropriate. However, in this work, we simply chose to
`cut-off' the estimate (Eq. \ref{Variance}) of the single-measurement variance before it reaches
zero at a value of $b_{i,\mbox{min}} = 0.04$.

\smallskip\ 

\subsection{ Distinguishing the `fingerprints' of different limit-cycles\label{chisquared}}

Since volatility (which we define below) requires an abundance of very {\em different} limit-cycles,
we need a highly-contrasting measure of whether a given limit-cycle is
different than or similar to another limit-cycle, or in essence
a method of distinguishing between limit-cycles' `fingerprints'. For each limit-cycle
(labelled by the slow time $t_s$), we measure each neuron's
fast-time-averaged firing rate, $A_i${\cal $(t_s)$} and its (cutoff)
fast-time variance (of the mean), $B_i${\cal $(t_s)$} (Eq. \ref{Variance2}),
and apply the weighted chi-square ($\chi ^2$) method
to decide whether two limit-cycles found at
different slow-times, $t_s$ and $t_s^{\prime }$, are similar or different:

\begin{equation}
\begin{array}{ccc}
\chi ^2(N,t_s,t_s^{\prime }) & = & \sum\limits_{i=1}^N\frac{
(A_i(t_s)-A_i(t_s^{\prime }))^2}{B_i(t_s)+B_i(t_s^{\prime })} 
\end{array}
\mbox{,} 
\end{equation}
where N is the number of neurons in the network as well as the
number of degrees of freedom for the $\chi ^2$-test. Each term in the $\chi
^2$ sum should approximate the square of a gaussian-distributed variable
with unit variance. The variance of the difference of two
gaussian-distributed variables is the sum of the individual variances (not
the average), as seen in the denominator of the $\chi ^2$ expression.

If the variance-estimate, $B_i(t_s)$, 
is appropriate and accurate, and if a class of similar
limit-cycles have gaussian-distributed time-averaged firing rates $A_i(t_s)$
, then we can estimate {\em a priori} the maximum value of $\chi
^2(N,t_s,t_s^{\prime })$ for similar limit-cycles. For an $N$-neuron
network, two similar limit-cycles should have (for $N>>1$) a chi-squared
given by: $\chi ^2(N,t_s,t_s^{\prime })\sim N \pm \sqrt{2 N}$.
Accordingly, in order to ensure that most of the 
similar limit-cycles are determined to be similar by the chi-square test, for
`similarity' we demand that $\chi ^2(N,t_s,t_s^{\prime })<N+3\sqrt{N}$. For
large $N$, chi-square tables indicate that this is approximately a $95\%$
confidence level experiment. This similarity test corresponds to
a tolerance of about 4 misfires per time step (for $\Delta W = 1000$)
From the $\chi ^2$-test, each pair of
limit-cycles is given a difference label $d(t_s,t_s^{\prime })$:

\begin{equation}
\label{difference}d(t_s,t_s^{\prime })=\left\{ 
\begin{array}{cc}
0\mbox{,} & \mbox{if }\chi ^2(N,t_s,t_s^{\prime })\leq N+3\sqrt{N} \\ 
1\mbox{,} & \mbox{if }\chi ^2(N,t_s,t_s^{\prime })>N+3\sqrt{N}
\end{array}
\right. \mbox{.}
\end{equation}
If $d(t_s,t_s^{\prime })=1$, $\forall t_s^{\prime }<t_s$, then the
limit-cycle found at slow-time step, $t_s$, is truly a {\em novel}
limit-cycle, never having been observed before. If the limit-cycle is novel
then it is given a `novelty' label of $n(t_s)=1$, otherwise $n(t_s)=0$ (by
default, the first observed limit-cycle will always be novel: $n(1)\equiv 1$). 
Simply by counting the number of slow-time steps in which a novel
limit-cycle is found, we can estimate the number of different limit-cycle
attractors, $N_d$, available to the RSANN:

\begin{equation}
\begin{array}{ccc}
N_d & \equiv & \sum\limits_{t_s=1}^{t_s^{\max }}n(t_s) \, . \\  
\end{array}
\end{equation}

\medskip\ 

\section{\bf The performance of the random asymmetric neural network}

\begin{figure}[p]
\centerline{\hbox{
\psfig{figure=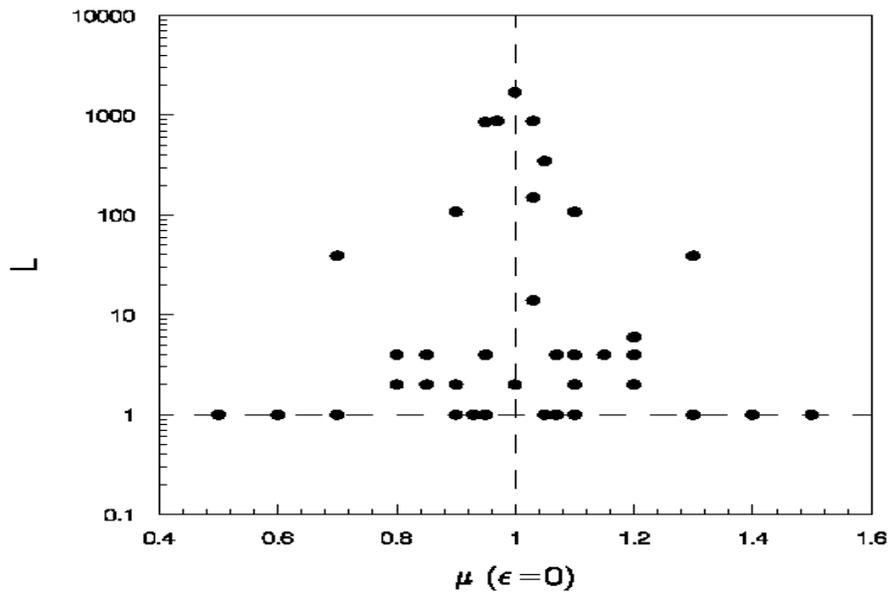,height=3.5in,width=15cm}
}}
\caption[Cycle Length vs. Bias Level]{The period, $L$, of observed limit-cycles 
as a function of the global-threshold-level $\mu$, for $\epsilon = 0$.
($N = 40$ neurons; also note the logarithmic L-scale).
\label{Lmu}}
\end{figure}

\begin{figure}[p]
\centerline{\hbox{
\psfig{figure=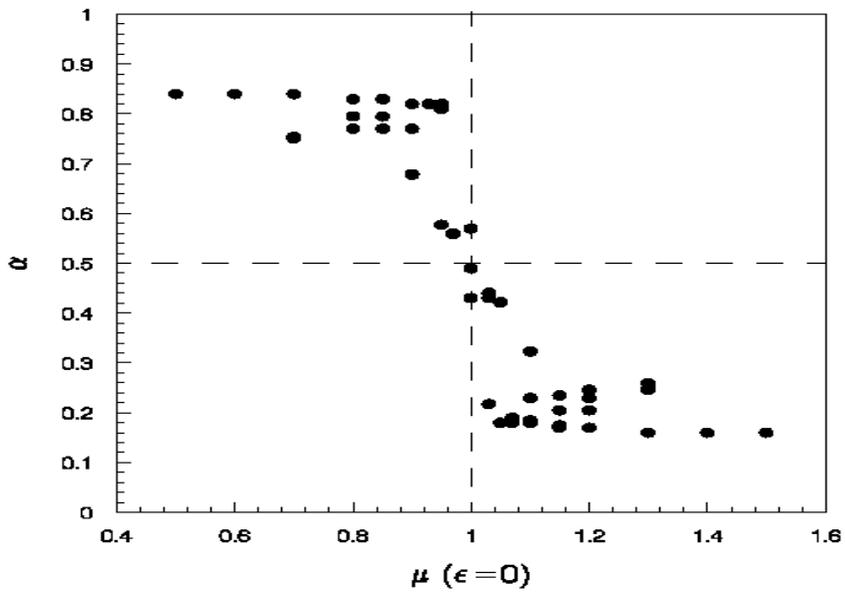,height=3.5in,width=15cm}
}}
\caption[Mean Activity vs. Bias Level]{The mean (over time and space) neural firing rate, $\alpha$,  for each observed
limit-cycle as a function of $\mu$ at $\epsilon=0$. 
\label{alphamu}}
\end{figure}

Random asymmetric neural networks will often develop into a fixed point
where the neural firing vector, $a_i(t),$ does not change in time, or in 
other words, the
limit-cycle has a period of one time step. For zero slow-threshold noise 
($\epsilon =0)$, if the mean threshold value is much greater(less) than
normal, then the RSANN will tend to have a fixed point with very few(many)
neurons firing each time step, which is called network death(epilepsy).
Likewise, for zero noise ($\epsilon =0)$ and for normal thresholds $(\mu =1)$, limit-cycles with very long periods are possible (Figs.\ref{Lmu},\ref
{alphamu}).

\begin{figure}[p]
\centerline{\hbox{
\psfig{figure=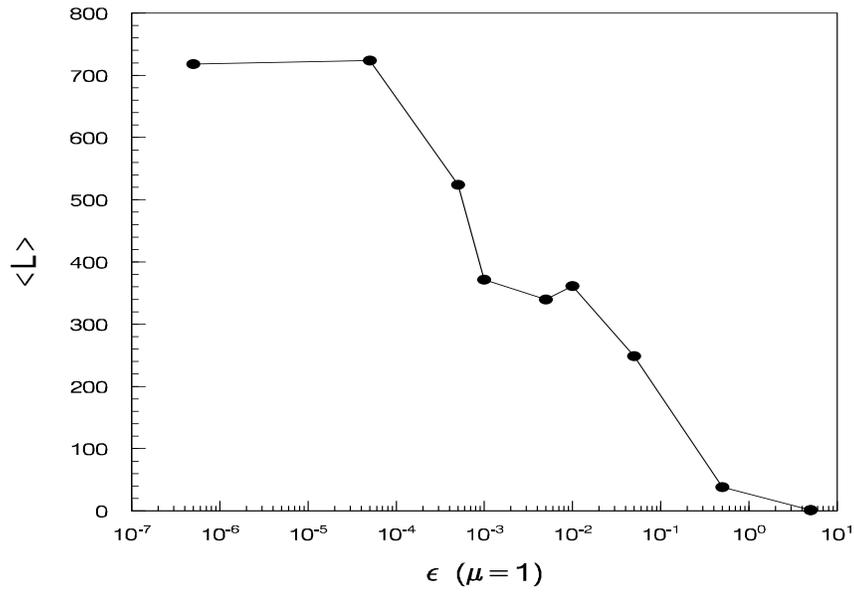,height=3.5in,width=15cm}
}}
\caption[Mean Cycle-Length vs. Slow-Noise Amplitude]{The period of the observed
limit-cycles, when averaged over slow-time for $t_{s,max} = 100$ slow-time-steps,
gradually decreases as the amplitude of the slow-noise increases.  
($\mu = 1$, $N = 40$ neurons;
note the logarithmic $\epsilon$-scale.
\label{LaveEps}}
\end{figure}

\begin{figure}[p]
\centerline{\hbox{
\psfig{figure=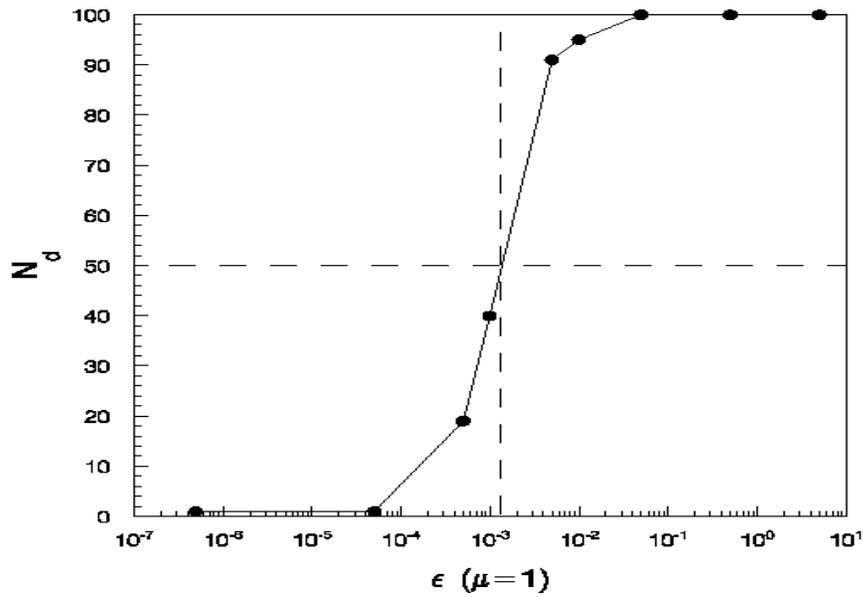,height=3.5in,width=15cm}
}}
\caption[Number of Different Cycles vs. Slow-Noise Amplitude]{The total number of
different limit-cycles observed (as determined from a $\chi^{2}$-comparison) quickly
approaches the total number of observations ($t_{s,max} = 100$), as the slow-noise
amplitude increases. The critical noise-amplitude is $\epsilon \approx 10^{-3}$.
\label{NdiffEps}}
\end{figure}

When the mean threshold value is normal ($\mu =1)$, but the spatiotemporal
threshold fluctuations from normality are large($\epsilon >\epsilon _2$),
then there also exist {\em many} different mixed death/epilepsy fixed points
in which a fraction of the neurons are firing at each time step and the
remaining neurons never fire. Conversely, when the mean threshold value is
normal and the spatiotemporal threshold fluctuations are small ($\epsilon
<\epsilon _1$), limit-cycles with very long periods are possible, but the
number of different limit-cycle attractors is limited. For normal mean
threshold values and intermediate-valued spatiotemporal threshold
fluctuations ($\epsilon _1<\epsilon <\epsilon _2),$ many intermediate-period
limit-cycles exist (Figs.\ref{LaveEps},\ref{NdiffEps}).

\smallskip\ 

\subsection{{\bf Eligibility as a function of noise-strength\label{elig}}}

\begin{figure}[htbp]
\centerline{\hbox{
\psfig{figure=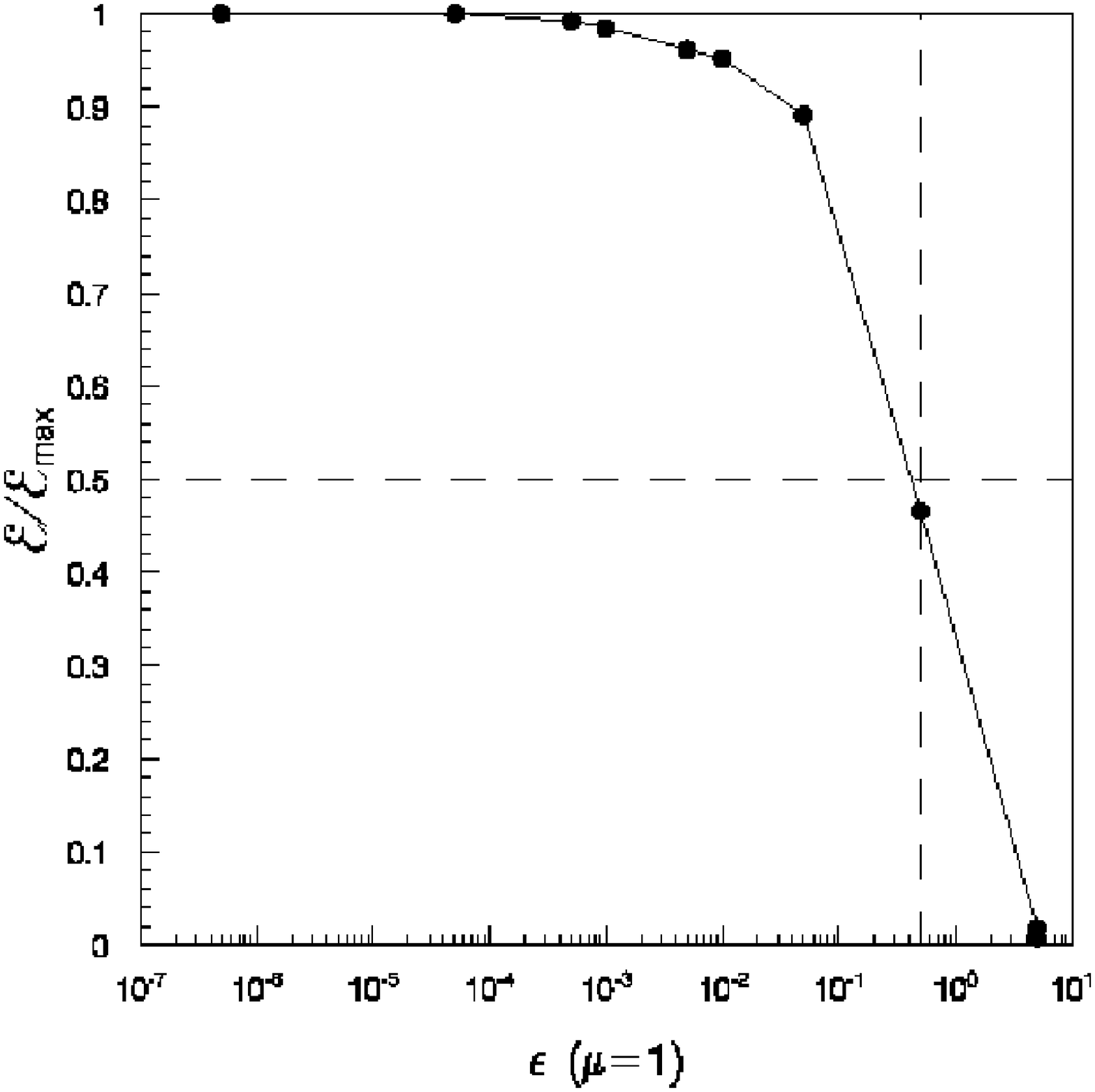,height=3.5in,width=15cm}
}}
\caption[Normalized Eligibility vs. Slow-Noise Amplitude]{The mean
eligibility ${\cal E}$ is shown as a function of the slow-noise
amplitude $\epsilon$, exhibiting a phase transformation at $\epsilon \sim 0.5$
to low eligibility..
\label{EEps}}
\end{figure}

A network is said to have a high degree of eligibility if many neurons
participate in the dynamical collective activity of the network. For an
attractor found at slow-time step $t_s$, the time-averaged spatial firing
vector, $A_i(t_s)=$$<a_i(t,t_s)>_t\in [0,1]$ (eq. \ref{activity}), will be
maximally eligible if $A_i(t_s)=0.5$ for all neurons $i,$ and minimally
eligible if $A_i(t_s)$$\in \{0,1\}$ for all neurons $i.$ The Shannon
information (or entropy) has these properties, so we will adopt the form of an entropy
function as our measure of the eligibility of each limit-cycle attractor, 
{\cal $e(t_s)$}:

\begin{equation}
e(t_s) \equiv -\frac 1N\sum_{i=1}^NA_i(t_s)\ln A_i(t_s)\,\,\,, 
\end{equation}
where the average eligibility, ${\cal E}$, per limit-cycle attractor is:

\begin{equation}
{\cal E}=\frac 1{t_s^{\max }}\sum_{t_s=1}^{t_s^{\max }}e(t_s) < \frac{1}{2}
\ln 2 = 0.3466 \equiv {\cal E}_{\max} \,\,\,. 
\end{equation}
Despite its utility, we do not have a detailed dynamical motivation for
using an entropy measure for our eligibility measure.
In Figure \ref{EEps}, we find that
eligibility goes through a phase transformation to zero at $\epsilon
_2\sim 0.3$, as the non-fixed-point limit-cycles gradually become fixed
points as $\epsilon $ increases. In other words, when the thresholds become
grossly `out-of-tune' with the mean membrane potential, the RSANN attractors
become more trivial, with each neuron tending towards its own independent
fixed point $a_i=1$ or $a_i=0$.

\smallskip\ 

\subsection{\bf Diversity}

We measure the accessibility of a given attractor by estimating the
probability, $P(t_s)$, that a given attractor (first seen at slow-time step
$t_s$, with the novelty label $n(t_s^{\prime})=1$ when $t_s^{\prime}=t_s$) is found during the slow-time observation period $t_s^{\prime }\in
[1,t_s^{\max }]$: 
\begin{equation}
\begin{array}{ccc}
P(t_s) & \equiv  & \frac{N(t_s)}{t_s^{\max }} \\ 
\end{array}
\,,
\end{equation}
where $N(t_s)$ is the total number of times the limit-cycle $t_s$ (first
observed at slow-time step $t_s$) is observed, and $n(t_s)$
is defined in Section \ref{chisquared}. Note that if
a given attractor is only observed {\em once} (at time step $t_s$), then $%
P(t_s)=1/t_s^{\max }$; also if the same attractor is observed at each
slow-time step, then $P(t_s=1)=1$.

\begin{figure}[htbp]
\centerline{\hbox{
\psfig{figure=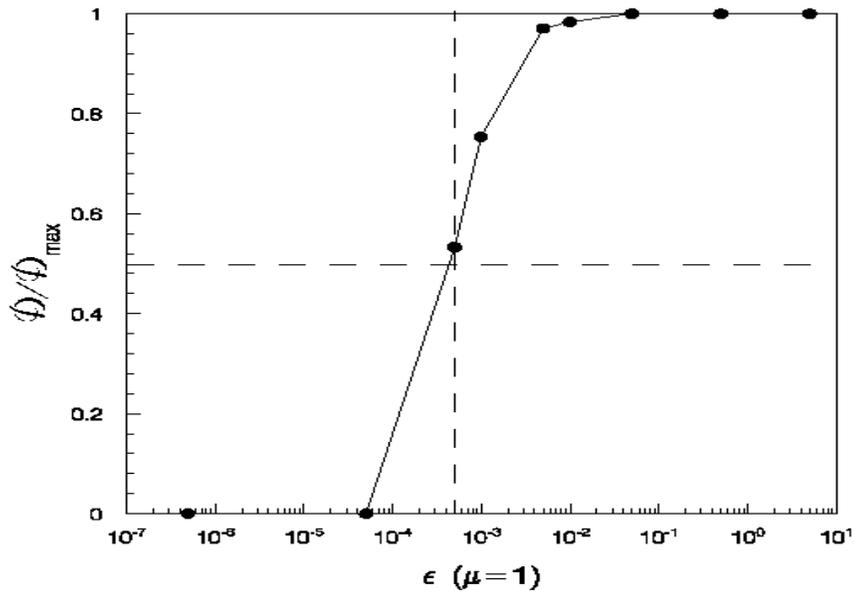,height=3.5in,width=15cm}
}}
\caption[Normalized Diversity vs. Slow-Noise Amplitude]{The diversity, an entropy-measure
of number of different limit-cycles, shows a phase transformation to maximal diversity at
a slow-noise amplitude $\epsilon \approx 0.0005 $ ($N = 40$ neurons). The maximum
diversity for $t_{s}^{max} = 100$ slow-time steps is ${\cal D}_{max} = \ln 100 \approx 4.6$.  
\label{DEps}}
\end{figure}

If each different limit-cycle attractor can be accessed by the network with
a measured probability, $P(t_s)$, we can define the diversity, ${\cal D}$,
as the inter-attractor occupation entropy: 
\begin{equation}
{\cal D}(t_s^{\max })=-\sum_{t_s=1}^{t_s^{\max }}P(t_s)\ln P(t_s) 
{\rm , \thinspace \thinspace \thinspace \thinspace \thinspace where}%
\sum_{t_s=1}^{t_s^{\max }}P(t_s)=1{\rm .} 
\end{equation}
It is easily seen that a large ${\cal D}$ corresponds to the ability to
occupy many different cyclic modes with equal probability; the diversity
will reach a maximum value of ${\cal D}={\cal D}_{\max }=\ln t_s^{\max },$
when $P(t_s^{\max })=1/t_s^{\max }$ for all slow-time steps $t_s^{\max }.$ A
small value for ${\cal D}$ corresponds to a genuine stability of the system
-- very few different cyclic modes are available. We observe a
phase transformation from low to high diversity by increasing the level of
randomness, $\epsilon $, on the threshold for each neuron past $\epsilon
_1\sim 5 \times 10^{-4}$ (Fig. \ref{DEps}). We have chosen not to explicitly divide
the diversity by $t_s^{\max }$ or $\ln t_s^{\max }$, since diversity should
be an extensive quantity, scaling with the observation time $t_s^{\max }$.
However, a diversity-production rate for the RSANN can easily be inferred, by
dividing by the observation time. We have found very little dependence of
the critical-point $\epsilon _1 \sim 5 \times 10^{-4}$ on the size of the network, $N.$

\smallskip\ 

\subsection{{\bf Volatility\label{volatile}}}

\begin{figure}[htbp]
\centerline{\hbox{
\psfig{figure=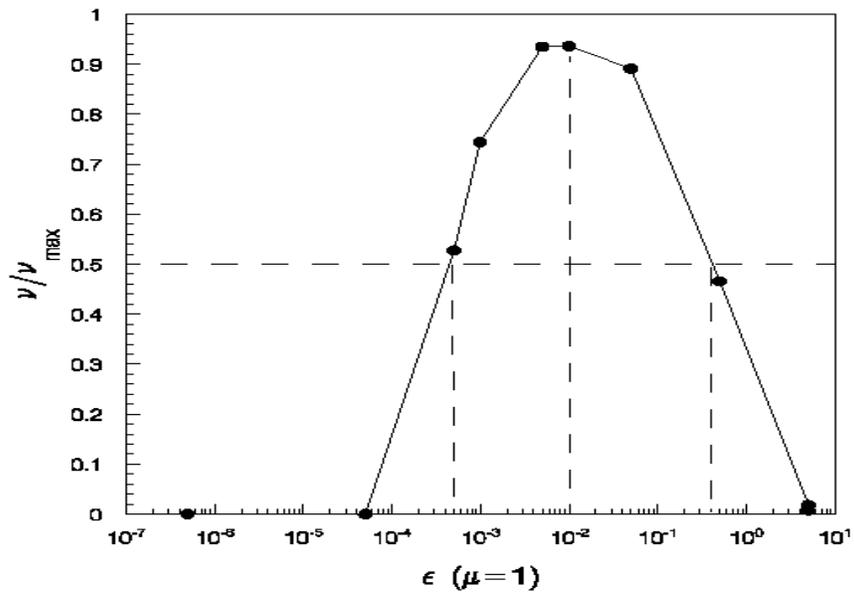,height=3.5in,width=15cm}
}}
\caption[Normalized Volatility vs. Slow-Noise Amplitude]{The volatility is large and
nearly maximal only within a broad range of slow-noise amplitudes: $5 \times 10^{-4} < \epsilon < 0.5$ 
($N = 40$ neurons, and $t_{s}^{max}= 100$). The maximal volatility for $t_{s}^{max} = 100$ slow-time
steps is ${\cal V}_{max} = \frac 12 \ln 2 \ln 100 = 1.5960$.
\label{VEps}}
\end{figure}

Volatility is defined as the ability to access a large number of
highly-eligible limit-cycles, or a mixture of high eligibility and high
diversity. We have defined eligibility and diversity above each in terms of
entropy-type measure. Therefore we shall define volatility also as
an entropy-weighted entropy: 
\begin{equation}
{\cal V}=+\sum_{t_s=1}^{t_s^{\max }}e(t_s)P(t_s)\ln P(t_s){\rm
\thinspace \thinspace .} 
\end{equation}
For ($\epsilon _1<\epsilon <\epsilon _2)$ (Fig. \ref{VEps}), volatility is
high, corresponding to two distinct transformations to volatility. At $%
\epsilon =\epsilon _1\sim 5 \times 10^{-4}$, the amplitude of slow threshold noise
causes a transformation to a diversity of different limit-cycles; at $\epsilon
=\epsilon _2\sim 5 \times 10^{-1}$, the slow threshold noise is so large that all
limit-cycles become fixed points. We accordingly label three different
regimes for the RSANN with slow threshold noise:

$\bullet $ Stable Regime: $\epsilon <5\times10^{-4}$

$\bullet $ Volatile Regime: $5 \times 10^{-4}\leq \epsilon < 5\times10^{-1}$

$\bullet $ Trivially Random Regime: $\epsilon \geq 5 \times 10^{-1}$.

\medskip 

\subsection{Net Size Dependence of Limit-Cycle Period and Number of
Different Attractors\label{netsize}}

\begin{figure}[htbp]
\centerline{\hbox{
\psfig{figure=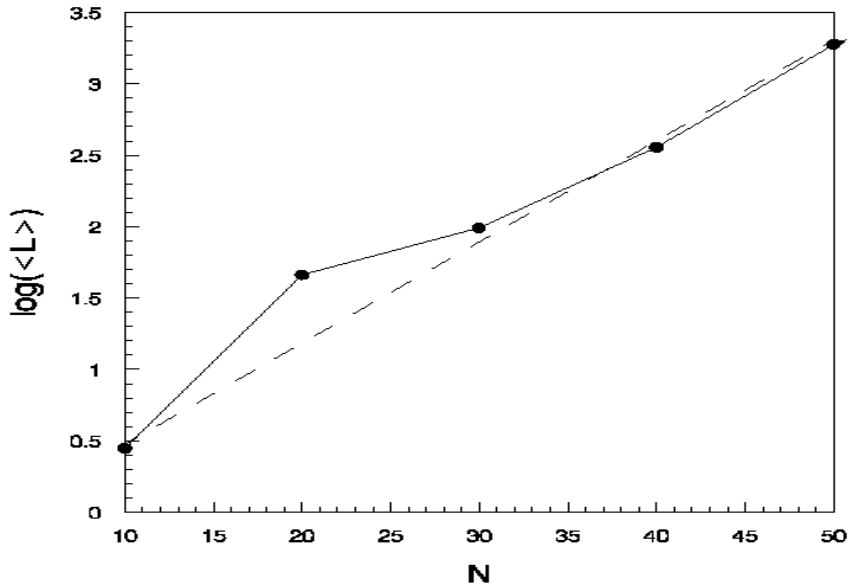,height=3.5in,width=15cm}
}}
\caption[Mean Cycle-Length vs. Number of Neurons]{The logarithm (base $10$) of
the mean-cycle length (within the volatile regime, $\epsilon = 0.01$) 
increases somewhat linearly with the size of the neural network, possibly 
implying that the volatile RSANNs are in the chaotic phase.
\label{logLaveN}}
\end{figure}

\begin{figure}[htbp]
\centerline{\hbox{
\psfig{figure=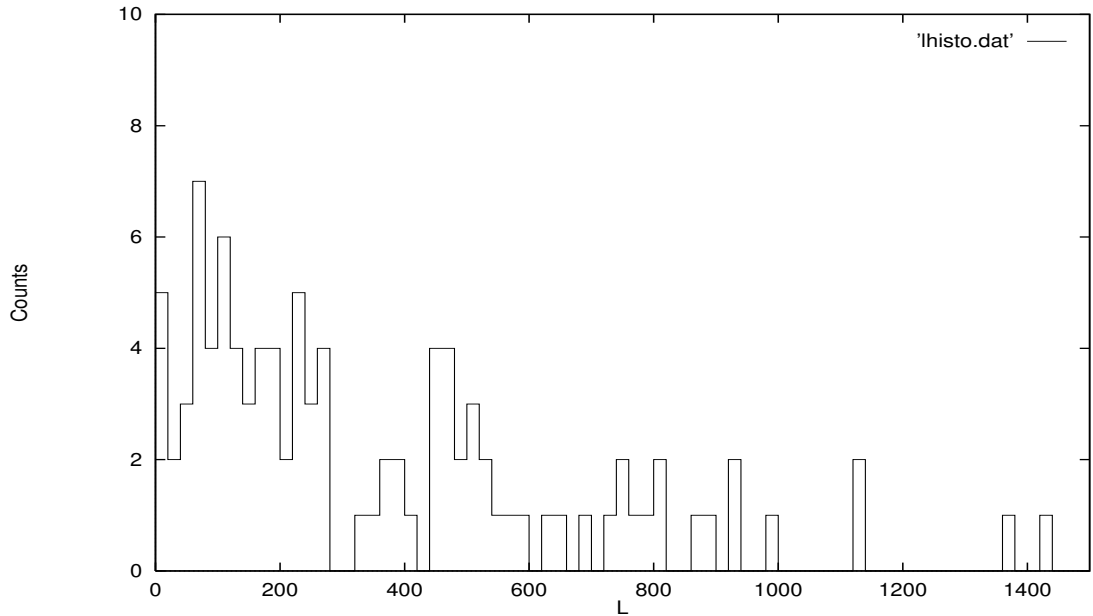,height=3.5in,width=15cm}
}}
\caption[Histogram of Limit-Cycle Periods]{
The distribution of different limit-cycle periods observed in the
volatile regime (for $N=40$, $P=2$ (see Section \ref{Updating}),
$\epsilon=0.01$).
\label{Lhisto}}
\end{figure}

From Table \ref{SizeTable}, for several different network sizes ($N\leq 50$)
with $M=10$ without slow threshold noise ($\epsilon =0$), we have found only
a handful of different limit-cycles ($N_{{\rm d}}\leq 6$) for each network,
without evidence for a systematic dependence on network size. The
limit-cycle periods for these networks at zero noise are dominated by one or
two different periods (typ($L$)), and also have no systematic dependence on
network size. We have found however that the transients (typ($W_{n-1}$))
prior to convergence to a limit-cycle tends grow very rapidly with network
size $N$. When $\epsilon =0.01$, which is in the volatile region, the
distribution of cycle lengths is broadly distributed (see Fig. \ref{Lhisto}%
), and from Table \ref{SizeTable}, the mean value $L_{{\rm ave}}$ of the
cycle length grows nearly exponentially with $N$ (Fig. \ref{logLaveN}). Since the
distribution of limit-cycle periods is highly non-gaussian (Fig.\ref{Lhisto}%
), caution should be used when interpreting the properties of the {\em %
average} cycle-length (as in Fig. \ref{logLaveN}). Potentially the maximum or
median observed cycle length should be used rather than the average.
Also, since the cycle length distribution does {\it not} exhibit peaks at regularly spaced intervals, the possibility of an errant limit-cycle comparison algorithm is unlikely.

\begin{table}[htbp]
\begin{center}
  \begin{tabular}{|l||l|l|l|l|} \hline
  &  $\epsilon=0$ & $\epsilon=0$ & $\epsilon=0$ & $\epsilon=0.01$ \\
     {\bf $N$}  &  {\bf $N_{d}$} 
 &  {\bf char($L$)} & {\bf char($W_{n-1}$)}
 &{\bf $<L>$} \\ \hline
     10       &  6       &  3     & 128   & 2.8 $\pm$ 0.06 \\
     20       &  3       &  142  & 512 & 46 $\pm$ 5 \\
     30       &  4       &  9     & 128    & 98 $\pm$ 9 \\ 
     40       &  1       &  688  & 2048  & 361 $\pm$ 32 \\
     50       &  5       &  2      & 4096  & 1886 $\pm$ 161 \\ \hline
  \end{tabular}
  \caption{For different network sizes, $N$, we tabulate the number, $N_d$, of different zero-noise limit-cycle attractors,
  characteristic cycle length (char($L(\epsilon=0)$), characteristic transient
  (char($W_{n-1}(\epsilon=0)$), and average cycle length with non-zero noise ($<L>(\epsilon=0.01)$).
  Each error bar is the R.M.S. variation from the mean divided
  by the square root of the number of initial-conditions. At each value of $\epsilon$, 
  we observe each net for $t_{s}^{max}=100$ slow-time steps, in which we require greater than
  $P=2$ periods (see Section 3.2) of limit-cycle repetition before searching for another
  limit-cycle.\label{SizeTable}}
\end{center}
\end{table}

\medskip\ 

\medskip

\medskip 

\subsection{Observation Period-Length Dependence of Number of Different
Attractors}

We observe in the volatile-regime, that the number of different attractors
observed, $N_d$, is nearly equal to the observation time-period $t_s^{\max }$,
when $\epsilon > 5 \times 10^{-3}$.
Conversely, in the stable-regime ($\epsilon < 5 \times 10^{-5}$), $N_d$ is largely independent of $%
t_s^{\max }.$ These two results are complementary: the former implying a
nearly inexhaustible source of different highly-eligible limit-cycle
attractors, the latter implying that we can access a small group of
different limit-cycle attractors with a high degree of predictability.
$N_d$ is often greater than $1$ (though small) for the
stable-regime, which means that the stable phase cannot be used to access a
particular attractor upon demand, but we can demand access to one of a small
number of different attractors. One might interpret
this result as showing that the RSANN can think the same thought in several
different ways, dependent on the initial conditions for the firing vector, $a_i(t=0,t_s)$.

\section{Conclusions and Summary}

\subsection{Comparison with other work}

K\"urten (1988a) has shown the existence of a dynamical phase
transition for zero-noise RSANNs with $K=3$ incoming connections per neuron
from a chaotic phase to a frozen phase. For $K<3$, RSANNs will always remain
in the frozen phase. For $K=3$, the chaotic-to-frozen transition can be
triggered either by randomly diluting the density of connections, or by
increasing an additive threshold parameter. In the chaotic phase, the RSANN
has sensitive dependence on initial conditions, and very long limit-cycle
trajectories (whose period scales as $L\propto 2^{\beta N}$). In the frozen
phase, the RSANN has insensitivity to initial conditions and very short
limit-cycle trajectories (whose period scales as $L\propto N^\alpha $). The
parameters $\alpha $ and $\beta $ depend on the connectivity $K$, the
additive threshold parameter, and the distribution of connection strengths.
Therefore, at zero threshold noise ($\epsilon =0$), our results are
inconclusive; however when we choose our thresholds from a sufficently wide
distribution ($\epsilon =0.01$), the limit-cycle period scales approximately
as an exponential function of $N$ (with an exponent of $\beta \sim 0.2$, see
section \ref{netsize}), so we might (cautiously) surmise that the volatile
RSANN is in the chaotic phase.

Kauffman (1993, pp. 191-235) and K\"urten (1988b)
shows that random boolean automata with $K=2$ (frozen phase) typically have $%
N_{d}\propto N^{1/2}$ different limit-cycle attractors (for each network
realization), whose period also scales as $L\propto N^{1/2}$. This small
number of different attractors is qualitatively consistent with our 
$\epsilon =0$ results. We need to perform an ensemble average of
many simulations to indeed confirm this result. We are currently unaware of any
definitive prior results on the number of different attractors for networks
in the chaotic phase (except for the observed preponderance of a very few
(long) limit-cycles, perhaps explained by `canalization' 
(Kauffman, 1993). However, the arguments of Derrida, Gardner, and Zippelius%
(1987) regarding the evolution of the overlap between a configuration
and a stored configuration might apply, which would imply a strict
theoretical upper limit on the number of storable/recallable limit-cycle
attractors at $N_d^{\max }=2N.$ However, we do not keep our thresholds
fixed, so this upper limit is not applicable, as seen in our results.

For our implementation of noisy RSANNs, we speculate that the number of
attractors in the chaotic phase scales exponentially with $N$. This
speculation is based on our observation that with our volatile RSANN, the
number of different attractors available at $\epsilon =0.01$ is nearly
inexhaustible since in the volatile phase we are always able to observe new
limit-cycle attractors just by giving the thresholds new initial conditions
from the narrow gaussian distribution.

\subsection{Volatility, Chaos, Stochastic Resonance, and Qualitative Non-Determinism}
Our original interpretation of $\epsilon >0$ as providing a slowly-varying
noise on the thresholds can be reinterpreted in a slightly different
perspective. The volatile RSANNs are probably in the chaotic phase even when $%
\epsilon =0$, but in order to observe the expected exponential dependence of
the cycle-length on $N$, we would need to perform an ensemble average (or
observe many different realizations of the RSANN from the same class with
different connection matrices). When $\epsilon >0$, we are actually sampling
a significant fraction of the entire ensemble. Noise can give a dynamical
system access to the whole ensemble of different behaviors at different
times during the lifetime of the dynamical system. Slowly-varying threshold
noise can act as a `scanner' for thoughts novel or long-lost.

Volatility can also be considered as a form of stochastic resonance,
in which an optimal noise amplitude enhances the
signal-to-noise ratio of a signal filter. Zero-noise or high-amplitude noise
tends to reduce the information-processing capability of a stochastic
resonant system. The observation of stochastic resonance without
external driving force (Gang {\it et al.}, 1993) has interesting parallels to
our observation of high volatility at intermediate noise amplitudes.

By taking advantage of the chaotic threshold parameters (the connection
strength parameters are probably also chaotic), we can access a large number
of different RSANN attractors. Hence with a feedback algorithm, one might be
able to construct a system to control this chaos (Ott, Grebogi \& Yorke, 1990)
and access a given attractor upon demand.
But this stability needs to be augmented by the ability to always
be able to access a novel attractor. This approach to controlled creativity
has been developed into the adaptive resonance formalism
(Carpenter \& Grossberg, 1987).
There is a theoretical proof (Amit, 1989, p. 66) that thresholds with
fast-noise (Little, 1974) (chosen from a zero-mean gaussian) and sharp
step-function non-linearities is equivalent to a system with zero threshold
noise but with a rounded `S'-curve non-linearity.
This proof does not apply to slowly varying threshold noise. Perhaps by choosing a {\em %
discontinuous} non-linearity, with additive noise in the argument of the
non-linearity, the volatility or creativity that we observe is due to {\em %
non}-determinism (H\"ubler, 1992), in which prediction of final limit-cycle
attractors is nearly impossible.
The notion of `qualitatively' uncertain dynamics (Heagy, Carroll \& Pecora, 1994)
describes non-determinism between different attractor basins.
Qualitative non-determinism differs from the effective quantitative
non-determinism observed in ordinary deterministic chaos, in that ordinary
chaos consists of a {\em single} `strange' attractor, while qualitative
non-determinism consists of {\em multiple} strange (or simple) attractors
`riddled' with holes for transitions to other attractors. Volatility is
precisely the same concept as qualitative non-determinism.

\subsection{Summary}

The main objective of this study has been to construct a volatile neural
network that could exhibit a large set of easily-accessible highly-eligible
limit-cycle attractors. Without noise, we demonstrate that random asymmetric
neural networks (RSANNs) can exhibit only a small number of different
limit-cycle attractors. With neuronal threshold noise within a rather wide
range ($\epsilon _1<\epsilon <\epsilon _2$), we show that RSANNs can access a
diversity of highly-eligible limit-cycle attractors. RSANNs exhibit a
diversity phase transformation from a small number of distinct
limit-cycle attractors to a large number at a noise amplitude of $\epsilon=
\epsilon _1\sim 10^{-4}$. Likewise, RSANNs exhibit a
eligibility phase transformation at a threshold noise amplitude of $\epsilon
=\epsilon _2\sim 0.5$.

\subsection*{Acknowledgements}

H. Bohr would like to thank P. Carruthers (now deceased), J. Rafelski,
and the U. Arizona Department of
Physics for hospitality during several visits when much of this work was
completed; P. McGuire thanks the Santa Fe Institute and A. H\"ubler and the
University of Illinois Center for Complex Systems Research for hospitality
and atmospheres for very fertile discussions. P. McGuire was partially
supported by an NSF/U.S. Dept. of Education/State of Arizona pre-doctoral
fellowship. We all thank many individuals who have provided different
perspectives to our work, including the following:
G. Sonnenberg, D. Harley, B. Skaggs,  
Z. Hasan, G. Littlewort, and J. Clark.

\end{document}